\documentclass[twocolumn,twoside,slac_two]{revtex4}
\usepackage{graphicx}
\usepackage{fancyhdr}
\begin{document}

%Title of paper
\title{The Gamma-Ray Blazar PKS 0208$-$512, a Multi-Wavelength Investigation}

% Repeat the \author .. \affiliation  etc. as needed
%
% \affiliation command applies to all authors since the last
% \affiliation command. The \affiliation command should follow the
% other information

\author{J. M. Blanchard, J. E. J. Lovell and J. Dickey}
\affiliation{School of Mathematics and Physics, Private Bag 37, University of Tasmania, Hobart TAS 7001, Australia }
\author{R. Ojha}
\affiliation{Astrophysics Science Division, NASA Goddard Space Flight Center, Greenbelt, MD 20771, USA}

\author{M. Kadler}
\affiliation{Institut fur Theoretische Physik und Astrophysik, Universitat Wurzburg, 97074 Wurzburg, Germany and CRESST/NASA Goddard Space Flight Center, Greenbelt, MD 20771, USA}

\author{R. Nesci}
\affiliation{University La Sapienza, Italy}

\author{P. G. Edwards and J. Stevens}
\affiliation{CSIRO Astronomy and Space Science, ATNF, Australia}

\author{M. Dutka}
\affiliation{The Catholic University of America, 620 Michigan Ave., N.E., Washington, DC 20064, USA}

\author{C. M\"{u}ller}
\affiliation{Erlangen Centre for Astroparticle Physics, Erwin-Rommel Str. 1, 91058 Erlangen, Germany}

\author{T. Pursimo}
\affiliation{Nordic Optical Telescope, Santa Cruz de La Palma, Spain}
\begin{abstract}
The gamma-ray blazar PKS 0208$-$512 has shown strong periods of flaring, at all frequencies from radio to gamma-ray. This has led to its inclusion in the TANAMI project, which tracks the jets of southern AGN using VLBI as well as supporting flux density monitoring programs. Time series analysis of the light curves generated by such monitoring is presented and discussed and VLBI maps of the source are used to show the evolution in the jet. A frequency dependent lag is observed between flaring at different radio frequencies which does not appear to correspond to purely optical depth effects. Major flaring at gamma-ray frequencies appears to be preceded by a new component in the jet seen in our VLBI data.

\end{abstract}

%\maketitle must follow title, authors, abstract
\maketitle

\thispagestyle{fancy}

% body of paper here - Use proper section commands
% References should be done using the \cite, \ref, and \label commands
% Put \label in argument of \section for cross-referencing
%\section{\label{}}

\section{Introduction}
 \begin{figure*}[p]
\centering
\includegraphics[width=0.9\textwidth]{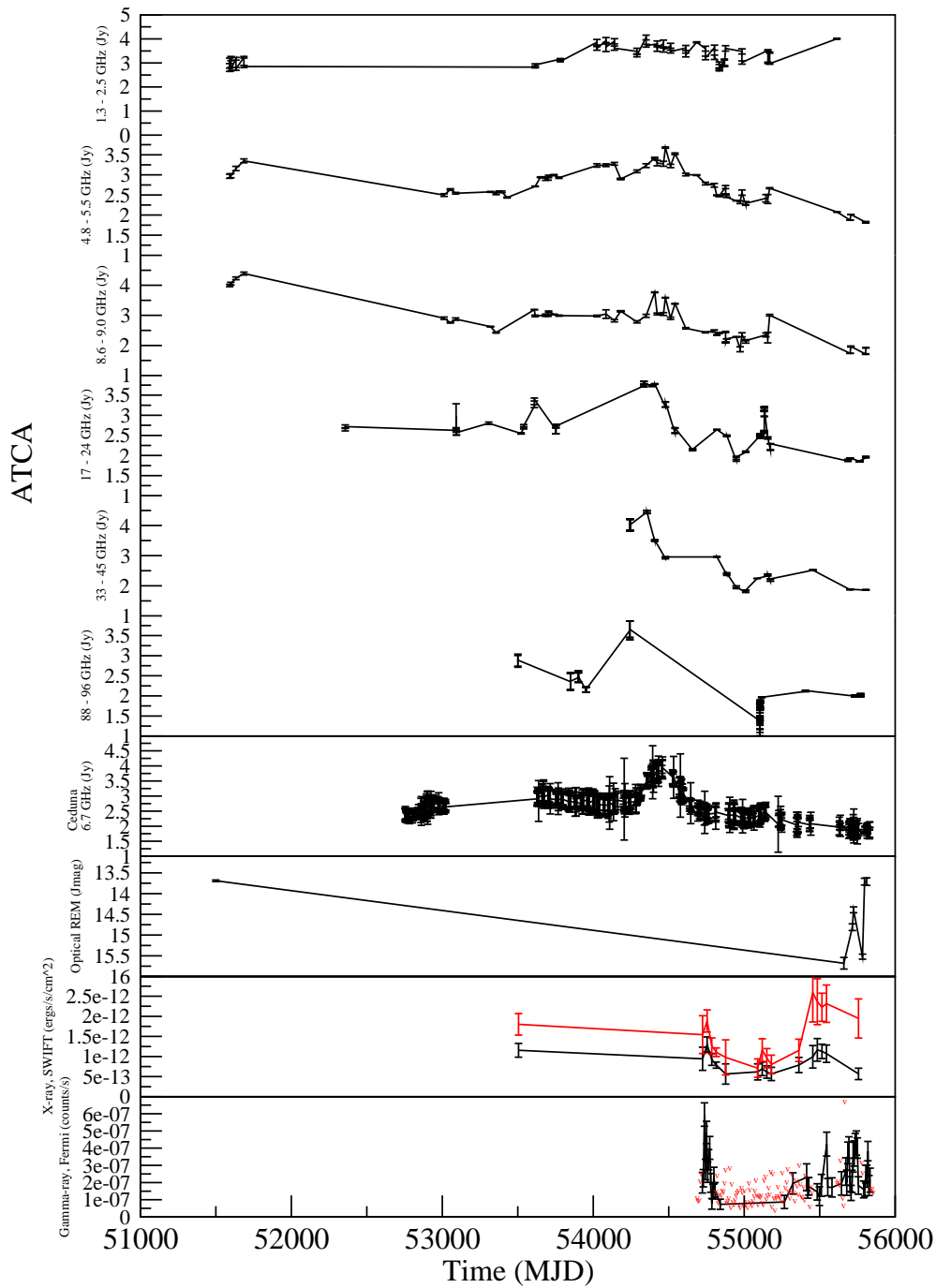}
\caption{The multiwavelength data assembled for PKS 0208$-$512. The X-ray SWIFT data is for the $0.2-2 keV$ (black) and $2-10 keV$ (red) range.}
\label{data}
\end{figure*}

\subsection{Justification}
The emission mechanism of gamma-rays observed from blazars is as yet poorly understood. It has been suggested that the mechanism itself could be inverse Compton upscattering of low energy photons \citep{Acciari2010}. These low energy photons could originate from within the jet itself, or from an external source such as broad-line regions, the accretion disk, or a dusty torus \citep{Sikora2009}. The gamma-ray emission is likely generated after the formation of a shocked region in the jet \citep{Lahteenmaki2003} so a delay between the radio and gamma ray emission generated by the event could be used to identify the location of the emission site. Early work in the field was done using the Energetic Gamma Ray Experiment Telescope (EGRET) \citep{Hartman1999}. EGRET detected gamma-rays in the 20 MeV - 30 GeV range and had a large effective field of view of roughly 30 degrees. It was however a pointed instrument and this meant that sources were typically monitored at fairly low cadence. Using EGRET data, \cite{Valtoaja1995}  suggested that the gamma-ray flares are preceded by high frequency radio flares, which would indicate a production mechanism involving strongly beamed gamma-rays from shocked regions of a jet. More recent work suggests that the gamma-ray emission is leading the radio (see for example \cite{Kovalev2009} or \cite{Pushkarev2010}). They suggest the emission is coming from the core of the blazar and that the delay is due to optical depth effects, giving a typical time lag of 1.2 months in the source frame.

Tracking Active Galactic Nuclei with Austral Milliarcsecond Interferometry (TANAMI) is a project to study the jets of sources south of -30 degrees declination using the Australian Long Baseline Array (LBA) \citep{Ojha2010}. This project carries out observations at 8.4 and 22 GHz at a cadence of roughly two months. The VLBI observations are typically made with all Australian LBA antennas (the Australia Telescope Compact Array (ATCA), Parkes, Mopra, Ceduna and Hobart) and often other southern hemisphere antennas including the Tidbinbilla 70m, TIGO, O'Higgins and Hartebeesthoek.

There are several facilities that are supporting these VLBI observations with varying monitoring projects. These include the ATCA, an interferometer consisting of six 22 m dishes, which monitors TANAMI sources at frequencies between 4.5 and 41 GHz every $\sim$6 weeks [Stevens et al. 2012, these proceedings]. Higher cadence monitoring is performed by the Ceduna 30 m antenna operated by the University of Tasmania which observes at frequencies between 2.4 and 22 GHz \citep{McCulloch2005}. Ceduna monitors approximately 40 TANAMI AGN at 6.7 GHz every two weeks, as well as higher cadence (daily) monitoring for a few sources which are also known to show short time scale (intra day) variability due to scintillation \citep{Senkbeil2009}.

\section{PKS 0208$-$512}

\begin{figure}[t]
\includegraphics[width=0.5\textwidth]{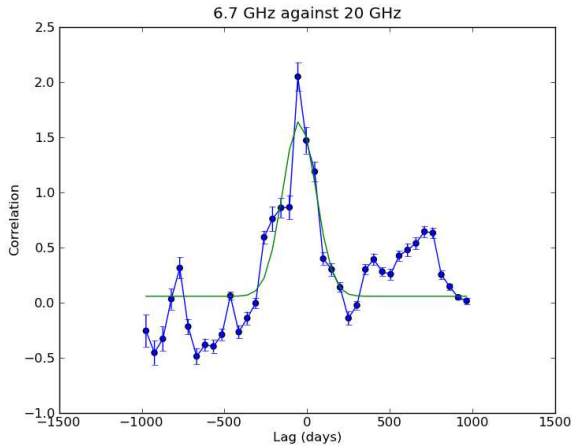}
\caption{An example discrete correlation function fit with a gaussian plus a flat baselevel using least mean squared. In this case this gives a correlation at a lag of $-47 \pm 16$ days between the 6.7 GHz Ceduna data and the 20 GHz ATCA data.}
\label{dcf}
\end{figure}
The blazar PKS 0208$-$512 is a redshift 0.999 AGN \citep{Wisotzki2000}. The source is classified as a blazar due to its high optical polarisation \citep{Scarpa1997}, flat radio spectrum, and variability at both optical and radio wavelengths. It shows a one sided jet at radio frequencies, extending to approximately 20 mas \citep{Tingay1996}. It is a strong and variable X-ray source, first detected by the R\"ontgen Satellite (ROSAT) \citep{Voges1999}. PKS 0208$-$512 was also detcted at gamma-ray frequencies by EGRET \citep{Bertsch1993}. The wealth of data available for this source, and the variability observed makes it an excellent target for an investigation into the link between the evolution of the radio jet and flaring at all frequencies.
\subsection{Data}
\begin{figure}[t]
\includegraphics[width=0.5\textwidth]{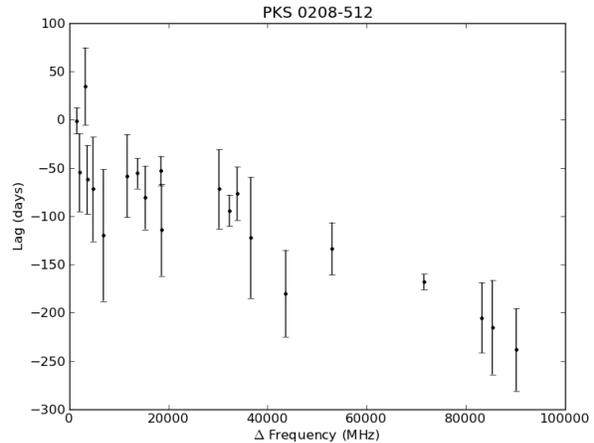}
\caption{The resultant lag in days for each radio frequency pair analysed using the DCF.}
\label{lag}
\end{figure}

PKS 0208$-$512 has been monitored at various frequencies for several years. This multi-wavelength data set is essential for the long term time series and cross correlation analysis we are attempting to perform. In particular, it is important to simultaneously sample as much of the spectral energy distribution (SED) as possible, allowing multi-epoch monitoring and timing/propogation of a flare through different wavelengths. Our data were obtained from the dedicated ATCA monitoring of TANAMI sources, as well as the ongoing ATCA calibrator monitoring program which operates between 1.1 and 96 GHz, and from Ceduna at 6.7 GHz, some optical data from the Rapid Eye Mount (REM) telescope, UV and X-ray data from SWIFT-XRT and gamma-ray data from the Fermi LAT Monitored Source List \citep{Atwood2009}. These data are presented in Figure \ref{data}.

\subsection{Discrete Correlation Function}

\begin{figure}[h]
\includegraphics[width=0.5\textwidth]{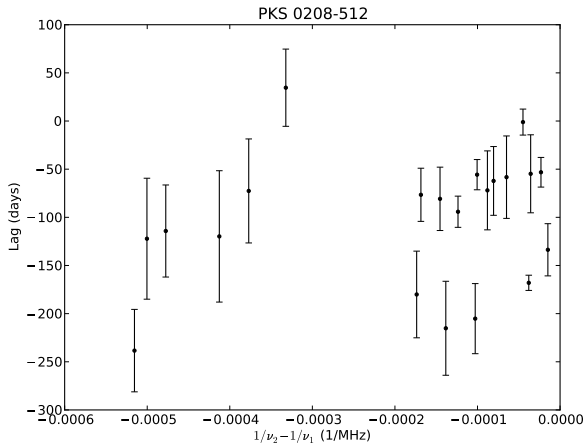}
\caption{The same as Figure \ref{lag}, except the x-axis has been changed to $1/\nu_2 - 1/\nu_1$. If optical depth effects are the sole cause of the time lag seen between radio pairs, a linear trend would be expected here, which is clearly not observed.}
\label{lag2}
\end{figure}

\begin{figure*}[t]
\centering
\includegraphics[width=0.7\textwidth,angle=-90]{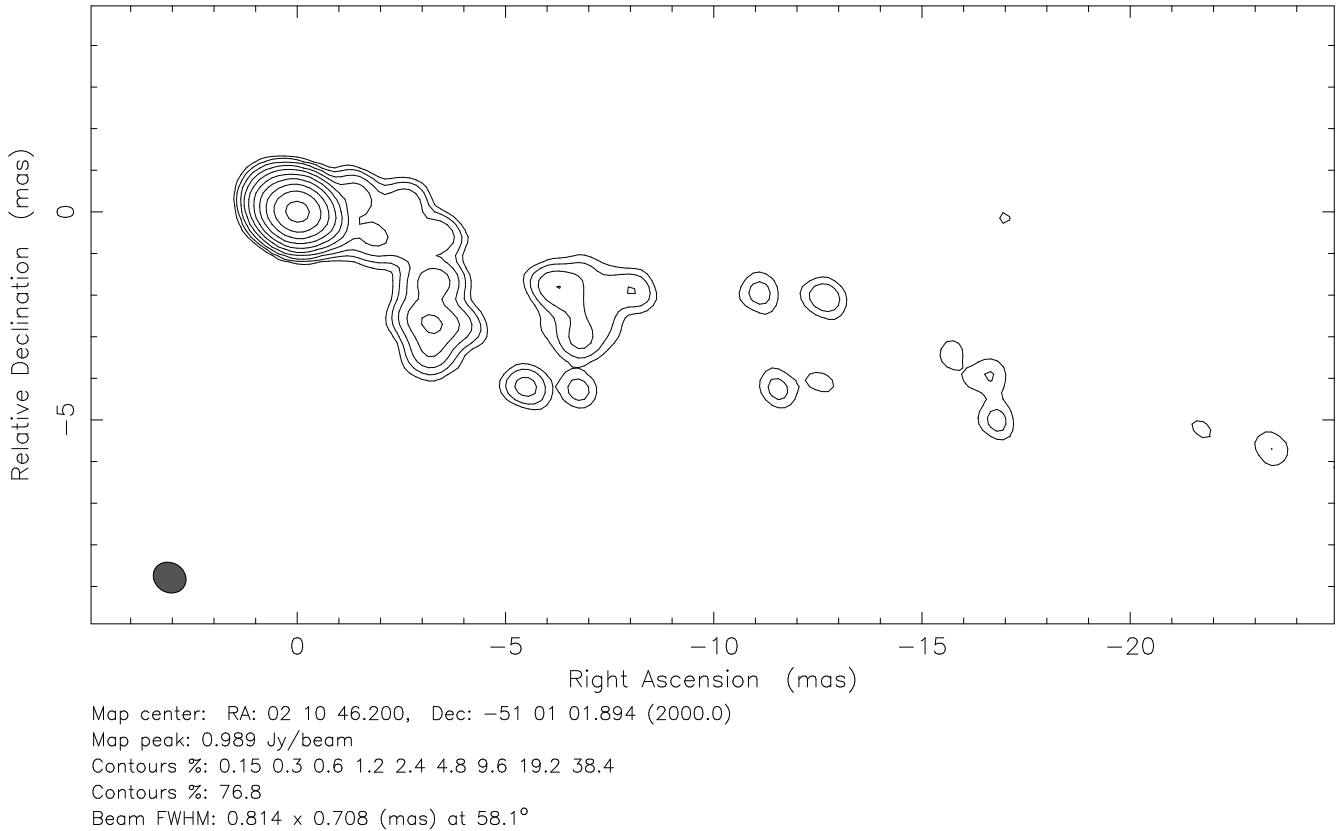}
\caption{VLBI image of PKS 0208$-$512. Note the one sided jet which bends strongly at around three mas from the core. }
\label{vlbi}
\end{figure*}

As previously mentioned, the time lag between flaring at various frequencies places limits on the emission regions, and hence the emission mechanisms occurring. In order to better understand the time lags seen between different radio frequencies, cross correlation of all the radio data pairs was performed using the discrete cross correlation function (DCF) described by \cite{Edelson1988}. The resultant correlation vs lag plot is then fitted with a gaussian using least mean squares techniques. The error in the fit is calculated from the covariance matrix obtained from the fitting algorithm. The discrete correlation function plots used here have not yet been normalized. This leads to correlation values of greater than one. In this preliminary analysis this has been ignored as the strength of the correlation is not being measured in this case, only the lag at which any correlation occurs. In the example shown in Figure \ref{dcf}, the gaussian fit to the DCF gives a lag of $-47 \pm 16$ days between the 6.7~GHz and 20~GHz lightcurves, i.e., with the 20~GHz variations leading the 6.7~GHz variations.

This was done for all radio frequency pairs and the resultant lags plotted against the difference in frequency. This is shown in Figure \ref{lag} and appears to show a clear correlation of increasing lag with difference between frequencies. At first glance this is not unexpected, \cite{Kovalev2009} already claimed that optical depth plays a major role in the delay between emission at different frequencies. However the nature of the realationship seen in Figure \ref{lag} is not consistent with purely optical depth effects. The critical radius at which the emission zone becomes optically thin ($R_c$) is related to frequency as shown in Equation \ref{od}.

\begin{equation}
R_c \propto \nu^{-1}
\label{od}
\end{equation}

 Applying this relationship to the lags calculated from the DCF should show a linear relationship between the lag and $\frac{1}{\nu_2} - \frac{1}{\nu_1}$. As seen in Figure \ref{lag2} this is not the case. Investigation of this effect is continuing.

\subsection{VLBI Observations}

\begin{figure}[t]
\centering
\includegraphics[height=0.5\textwidth,angle=-90]{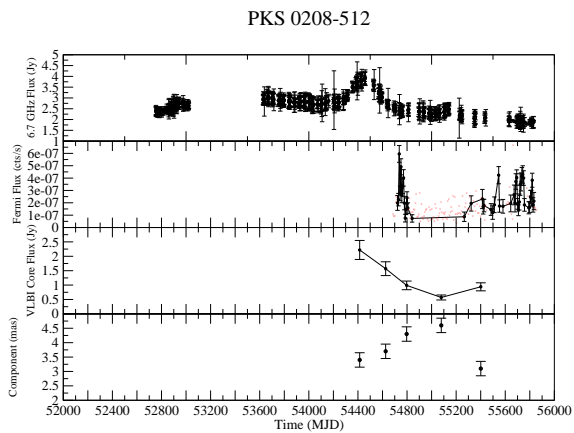}
\caption{The VLBI core flux density and first component separation from the core compared with the Ceduna radio and the Fermi gamma-ray light curve.}
\label{mod}
\end{figure}

PKS 0208$-$512 has been observed roughly once every six months since the end of 2007 as part of the TANAMI campaign of observations. The source shows a one sided jet, extending several mas, with a sharp bend roughly three mas from the core (see Figure \ref{vlbi}). Five epochs of the VLBI observations have been reduced to date, allowing for a basic kinematic study to commence. Basic modeling using DIFMAP \citep{Shepherd1994}  of the first discrete jet component out from the core has been done, and the distance of this component from the core, as well as the core flux density is shown in Figure \ref{mod}. The core flux density agrees well with the single dish monitoring flux density at 6.7 GHz from the Ceduna telescope. The first jet component separation from the core appears to be increasing at around 0.5 c. In the fifth epoch a much closer separation was measured, suggesting a new component being generated. This occurs just before the next major flaring event seen by Fermi and is consistent with the theory that gamma-ray emission occurs due to new shocked regions in the jet. An alternative theory is well described by \cite{Lister2009}. The new component may actually be emerging much closer to the core, pulling the core centroid toward the jet component and making it appear as if the separation has reduced. This would still imply a new component being generated in the core, as expected due to the gamma-ray flare.

\section{Further Work}

The discrete cross-correlation function analysis is being extended to include optical/UV, X-ray and gamma-ray data, to examine any possible correlation between these frequencies. Several more VLBI epochs have been observed and are being calibrated and imaged. This, along with more detailed modeling of the jet components, will give insight into the relationship between new jet components observed at radio frequencies and flaring seen at higher energies. 

The associated SWIFT-XRT UV data will provide another lightcurve at an intermediate frequency between radio and gamma-ray, allowing further analysis of any time lag vs frequency dependence observed. The UV data will also be crucial for the construction of SEDs currently being calculated for this source during both quiescent and flaring periods in the gamma-ray lightcurve. Examining the evolution of the SED of PKS 0208$-$512 during a gamma-ray flaring event gives a glimpse into the physical processes governing energy generation and emission occuring.
1

\bigskip

\begin{acknowledgments}
This research was funded in part by NASA through Fermi Guest Investigator grants NNH09ZDA001N and NNH10ZDA001N. This research was supported by an appointment to the NASA Postdoctoral Program at the Goddard Space Flight Center, administered by Oak Ridge Associated Universities through a contract with NASA.
\end{acknowledgments}

\bigskip % extra skip inserted
% Create the reference section using BibTeX:
\bibliography{ref}
%\begin{thebibliography}{9}   % Use for  1-9  references
%\begin{thebibliography}{99} % Use for 10-99 references

%\bibitem{accelconf-ref}
%http://www.cern.ch/accelconf

%\bibitem{exampl-ref}
%A.N. Other, ``A Very Interesting Paper'', EPAC'96, Sitges, June
%1996.

%\bibitem{templates-ref}
%http://www.cern.ch/accelconf/templates.html

%\end{thebibliography}

\end{document}